\documentclass[twocolumn,prl,aps]{revtex4}

\usepackage{graphicx}
\usepackage{dcolumn}
\usepackage{bm}

\begin{document}

\preprint{APS/123-QED}

\title{Two regimes in the magnetic field response of superconducting MgB$_2$}

\author{A. Kohen*, F. Giubileo , Th. Proslier*, F. Bobba, Y. Noat*, A. Troianovski*,
A. M. Cucolo, W. Sacks*, J. Klein*, D. Roditchev*}
\affiliation{%
Physics Dept. and INFM-SUPERMAT Lab., University of Salerno, via
S. Allende, 84081 Baronissi (SA), Italy
\\
*Groupe de Physique des Solides, Universit\'es Paris 6 et 7,
C.N.R.S. (UMR 75\ 88),  75015 Paris, France
}%

\author{N. Zhigadlo, S.M. Kazakov, J. Karpinski}
\affiliation{ Solid State Physics Laboratory, ETH Zurich, CH-8093
Zurich, Switzerland
}%

\date{\today}

\begin{abstract}
Using Scanning Tunneling Microscope at low temperature we explore
the superconducting phase diagram in the $\pi$-band of the
two-band superconductor MgB$_2$. In this band the peculiar shape
of the local tunneling spectra and their dynamics in the magnetic
field reveal the complex character of the quasiparticle density of
states (DOS). The gap in the DOS is first rapidly filled with
states in raising the magnetic field up to 0.5 T and then slowly
approaches the normal state value : The gap is observed up to 2 T.
Such a change in the DOS dynamics suggests the existence of two
terms in the DOS of the $\pi$-band: a  first one, reflecting an
intrinsic superconductivity in the band and a second one,
originating from an inter-band coupling to the $\sigma$-band. Our
findings allow a deeper understanding of the unique phase diagram
of MgB$_2$.
\end{abstract}

\pacs{74.50.+r 
74.70.Ad 
74.25.Ha 
                  }           
\maketitle

The discovery in January 2001 of superconductivity at 39K in the
well-known binary compound MgB$_2$ \cite{Akimitsu} has not only
opened a new way for potential applications of superconductors but
also offered to the scientific community an unexpected field of
fundamental research. A huge effort was very soon rewarded:
specific heat \cite{Budko}, tunneling
\cite{Giubileo1,Giubileo2,Iavarone} and point-contact spectroscopy
(PCS) \cite{Szabo,Gonnelli} experiments along with the theoretical
predictions \cite{Kortus,Liu} have shown MgB$_2$ to be a very
unusual superconductor in which two electronic bands contribute to
the superconductivity in a different way. It was established that
in the superconducting state two gaps open at the Fermi level: a
leading one, of about $\sim$6-7 meV in the two-dimensional
$\sigma$-band and a weak one, of $\sim$2-3 meV in the
three-dimensional $\pi$-band.

Since then, the effect was extensively studied in a large number
of experimental and theoretical works (for a review see
\cite{PhysC}). It was shown that the superconducting gaps are
strongly coupled: Both close at the same critical temperature of
39K \cite{Giubileo2}. Thus, in MgB$_2$ superconductivity is not
described by two independent condensates of Cooper pairs but
rather by a single condensate of a complex nature.


Various studies revealed a peculiar magnetic field response of
MgB$_2$. The study of the vortex lattice by Scanning Tunneling
Microscopy/Spectroscopy (STM/STS) \cite{Eskidsen} evidenced an
unusually large size of the $\pi$-band vortex cores. This
discovery led the authors to suggest that gap in the $\pi$-band is
induced by the intrinsically superconducting $\sigma$ band. Heat
capacity measurements have shown an anisotropy for H$\parallel c$
and H$\parallel ab$ \cite{heat}. Remarkably, this anisotropy
appears only for H$> 0.5$ Tesla, with a large anisotropy
developing at higher fields and leading finally to H$_{c2} \sim$3
T (H$\parallel c$) and $\sim$20 T (H$ \parallel ab$). Neutron
diffraction experiments revealed a rotation of the flux lattice by
30$^{\circ}$ in a field of $\sim 0.6-0.8$ T, occurring
simultaneously with a sharp drop in the diffraction peak intensity
\cite{NMR}. These results were attributed to the suppression of
the superconductivity in  the $\pi$-band at low fields.
Experimentally, the magnetic field response of the two bands was
studied in numerous PCS reports
\cite{Szabo,Gonnelli,Szabo1,Gonnelli1}, however the $\pi$-band gap
could be evaluated with reasonable accuracy only  up to 0.5 T.
Theoretical works regarding the mixed state in a two-band
superconductor \cite{Nakai,Koshelev,Dahm} have qualitatively
explained the above findings as a result of the inter-band
coupling. Finally, though many experiments revealed a modification
in the superconducting properties at the magnetic field of around
0.5-0.8 T, none of the observed changes were unambiguously linked
to the evolution of the principal superconducting parameters such
as the superconducting gap(s) and/or the inter-band coupling.

In this Letter we address the question of the magnetic field
response of the $\pi$-band by means of tunneling spectroscopy.
Using STM/STS we studied c-axis oriented surface of MgB$_2$ single
crystals.  We found the $\pi$-band DOS to deviate significantly
from the BCS shape even at 0 T, and to have a non-trivial field
dynamics. We observed two distinct regimes: A rapid evolution at
low fields B $\lesssim$ 0.4 T and a slow one at high fields B
$\gtrsim$ 0.6 T, separated by a crossover region where the DOS
remains almost unchanged. A clear gap in the $\pi$-band was still
observed at 2 T. Such a dynamics suggests the existence of two
terms in the $\pi$-band DOS: A first one, reflecting the
contribution to the superconducting gap from the electron-electron
interaction via phonons within the $\pi$-band, and a second one,
originating from an inter-band coupling to the $\sigma$-band.
These two terms behave differently in the magnetic field, the
first term almost vanishing in the field of 0.6 T. Our
experimental findings show the need for an additional theoretical
effort in the case of superconductivity in MgB$_2$.

Single crystals of MgB$_2$ were grown by a high pressure method
described elsewhere \cite{Karpinski}. The experiments were carried
out on a low temperature Omicron STM at Salerno University and on
a home-built STM/STS in Paris. The tunneling junctions were
achieved by approaching mechanically cut Pt/Ir tips to the c-axis
oriented surface of the crystals. Due to the peculiar electronic
structure of the material, c-axis tunnelling allows to probe
mainly the quasiparticle density of states (DOS) in the $\pi$-band
\cite{Brinkman,Eskidsen}. In such a geometry the $\pi$-band
component of the local DOS was studied in magnetic field up to 3T.
The superconducting critical temperature was determined locally by
measuring the evolution of the tunneling conductance spectra as a
function of temperature, and was found to be 38.5K.

The field dependence of the tunneling conductance spectra
$dI/dV(V)$ in the $\pi$-band was obtained by fixing the STM tip in
a selected location of the sample and by continuously measuring
local $I-V$ tunnelling curves while sweeping the magnetic field,
the field in all measurements being parallel to the c-axis. The
samples were cooled in zero magnetic field. At low temperature,
the field was slowly increased from zero up to the maximum value
and reduced to zero. Following, the same procedure was repeated
for the reversed polarity. The field sweep rate was
$\sim0.01-0.05$ T/min and the acquisition time for a single
spectrum was $\sim$10 msec. During such a short acquisition time
the field change of $\Delta B \simeq 10^{-6}$ T is, much smaller
than the field magnitude B itself, $\Delta B / B << 1$. Thus, the
field may be considered as constant for every single data set. In
type II superconductors however, the field sweep induces vortex
entry into the sample, which modifies the spectra. Indeed, while
the STM tip remains fixed the vortices move and from time to time
appear under the tip. Statistically some tunneling spectra reflect
the DOS in the vortex core, others correspond the DOS between the
vortices. Averaging over a large (10$^2$-10$^3$) number of
consecutively acquired spectra results in a smooth curve
representing the quasiparticle DOS spatially averaged over the
vortex unit cell \cite{Note}.
\begin{figure}
\includegraphics[width=7.7cm]{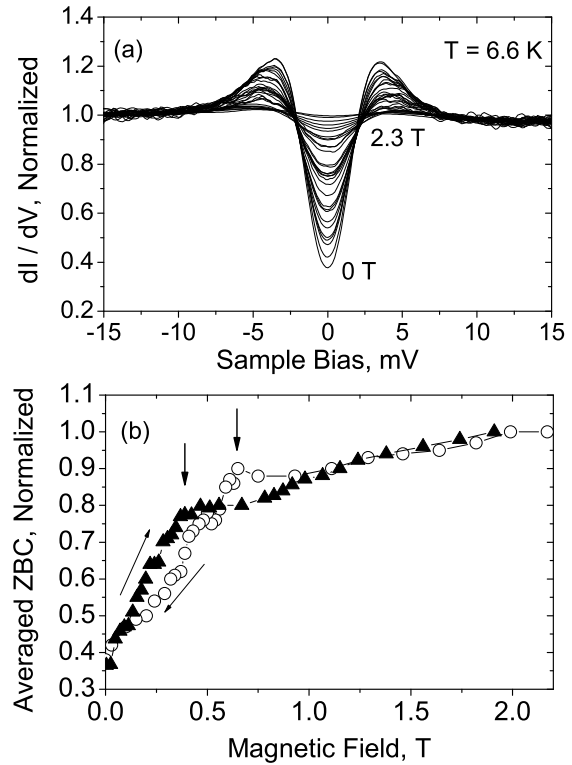}
\caption{\label{fig:epsart} (a) Evolution of the normalized
tunneling conductance spectra in the magnetic fields up to 2.3
Tesla at T = 6.6 K. (b) Evolution of the Zero Bias Conductance in
increasing (triangles) and decreasing (circles) magnetic field.}
\end{figure}
In Fig.1a we show the dynamics of the tunneling conductance
spectra (averaged over the vortex lattice unit cell) as a function
of the magnetic field. The spectra are characterized by a gap near
zero-bias and by only two of the four coherence peaks at $\pm$ 3
meV, consistent with what is expected for the tunneling to the
$\pi$-band and in agreement with previous experimental data
\cite{Giubileo1,Szabo,Gonnelli,Iavarone}.
The main effect of the magnetic field on the tunneling DOS in Fig.
1a is the filling with the states inside the gap. The field
dynamics of the DOS at the Fermi level is represented by the
Zero-Bias Conductance (ZBC) in the tunneling spectra (Fig.1b). The
evolution of the ZBC in the magnetic field is unusual : At low
fields, the ZBC rises rapidly and reaches a value of about 80\% of
the normal state ZBC already at 0.4 T. At higher fields however,
the dynamics drastically changes: the filling of states becomes
much more slower and, even at 2.0T, it is still possible to
distinguish the quasiparticle peaks and a minimum in between in
the raw tunneling conductance spectra. At intermediate fields (0.4
$< B <$ 0.7 T) there is a crossover region in which the ZBC
remains roughly constant. Remarkably, this range matches exactly
the range in which a rotation of the vortex lattice was observed
\cite{NMR}. By lowering the field we observe a similar dynamics,
though the crossover region is slightly shifted to higher fields.
Such a slightly hysteretic behavior indicates different vortex
dynamics in increasing and decreasing fields, which may be due to
geometrical barriers, vortex pinning, and lattice re-arrangements.
\begin{figure}
\includegraphics[width=8.7cm]{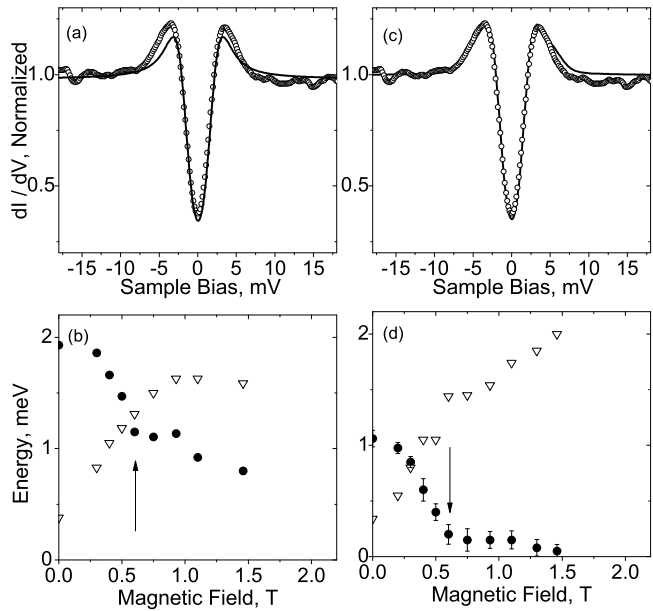}
\caption{(a) Zero field spectrum fitted by BCS model (1): T= 7 K,
$\Delta$=1.96 meV and $\Gamma_{Dynes}$= 0.32 meV. (b) Magnetic
field dynamics of superconducting gap (black circles) and pair
breaking parameter $\Gamma_{Dynes}$ (open triangles) extracted
from BCS fit of the tunneling spectra. (c) Zero field spectrum
fitted using model (2): T= 6.6 K, $\Delta_{\pi}^0$ = 1.1 meV,
$\Delta_{\sigma}^0$ = 6.0 meV, $\Gamma_{\pi}$=1.3 meV,
$\Gamma_{\sigma}$= 0.9 meV and $\Gamma_{Dynes}$ = 0.3 meV. (d)
Magnetic field dynamics of superconducting gap (black circles) and
pair breaking parameter $\Gamma_{Dynes}$ (open triangles)
extracted from model (2).}
\end{figure}
In order to follow such an unusual field dynamics we use two
different models describing the DOS in the $\pi$-band. The first
one considers the superconducting gap to be composed of two terms:
a first one originating from the electron-electron interaction via
phonons inside the $\pi$-band, and a second one, arising from the
superconducting coupling to the $\sigma$-band via phonon exchange.
The resulting superconducting gap $\Delta_{\pi}$ is calculated
self-consistently \cite{Koshelev}, but finally, the DOS in the
$\pi$-band remains BCS-like:
\begin{equation}
N_{\pi}(E,\Gamma_{Dynes})\propto
Re\left[\frac{|E|-i\Gamma_{Dynes}}{\sqrt{(|E|-i\Gamma_{Dynes})^2-\Delta_{\pi}^{2}}}\right]
\end{equation}
In this Dynes formula \cite{Dynes} the pair-breaking parameter
$\Gamma$ accounts phenomenologically for the effect of the
magnetic field on the DOS.

The second model, in addition, considers the quasiparticle
inter-band scattering. To analyze the shape of the DOS in such a
case we use, as a first step, two coupled equations proposed by
McMillan \cite{MacMillan} for the proximity effect in real space.
This formalism was already successfully applied in some previous
works \cite{Chinese,Schmidt}, and it's basic idea is well
described in \cite{Schmidt}. In the framework of this approach,
the DOS in each band deviates from the BCS shape as both
$\Delta_{\pi}$ and $\Delta_{\sigma}$ become energy dependent and
the apparent gap in the quasiparticle spectrum of the $\pi$-band
is larger than the self-consistently calculated superconducting
gap, due to the quasiparticle scattering from the $\sigma$-band:
\begin{equation}
\Delta _\pi  \left( E \right) = \frac{{\Delta _\pi ^0  +
\frac{{\Gamma _\pi  \Delta _\sigma  \left( E \right)}}{{\sqrt
{\left( {\Delta _\sigma  \left( E \right)} \right)^2  - E^2 }
}}}}{{1 + \frac{{\Gamma _\pi  }}{{\sqrt {\left( {\Delta _\sigma
\left( E \right)} \right)^2  - E^2 } }}}}
\end{equation}
\[
\Delta _\sigma  \left( E \right) = \frac{{\Delta _\sigma ^0  +
\frac{{\Gamma _\sigma  \Delta _\pi  \left( E \right)}}{{\sqrt
{\left( {\Delta _\pi  \left( E \right)} \right)^2  - E^2 } }}}}{{1
+ \frac{{\Gamma _\sigma  }}{{\sqrt {\left( {\Delta _\pi  \left( E
\right)} \right)^2  - E^2 } }}}}
\]
The $\pi$-band DOS is still given by Eq. (1), however
$\Delta_{\pi}$ is replaced by $\Delta_{\pi}(E)$, where$\Delta _\pi
^0$ and $\Delta _\sigma ^0$ represent the intrinsic pairing
potentials. The effect of finite temperature is accounted for in
both models in a standard way by the convolution integral of the
DOS (1) with the derivative of the Fermi-Dirac function.

We now focus on the dynamics of the superconducting gap in the
magnetic field using the above models. As we will show in both
cases the superconducting gap $\Delta_{\pi}$ exhibits a strong
change near 0.6T (for the decreasing field branch). In Fig.2a we
present an example of a BCS fit (1) to the raw tunneling zero
field conductance spectrum. We note that this fit fails to
reproduce simultaneously the value of the ZBC and the amplitude of
the quasiparticle peaks for any field. However, it does allow an
estimate for the energy scale of the gap. Fig.2b shows the field
dynamics of the main parameters extracted from the BCS fits. In
particular, one can see that the superconducting gap exhibits a
rapid drop at 0.6T while the pair-breaking rises continuously in
this field region. In Fig.2c we show a typical fit of the same DOS
as in Fig.2a using McMillan model (2). Within this model we find a
much better agreement between the experimental data and
theoretical curves. Also in this case, the superconducting gap
extracted from the fits exhibits a remarkably strong damping at a
field around 0.6T (black circles in Fig.2). In all calculations
$\Gamma_{\pi}$ and $\Gamma_{\sigma}$  are kept constant since
physically the inter band scattering rates should not change
significantly in the magnetic field.


The use of McMillan model is not arbitrary but motivated by the
fact that experimentally we observe systematic deviations of the
tunneling spectra from BCS behaviour. Such discrepancies are
better seen in SIS spectra as shown in Fig.3. The spectra were
measured at zero field using a superconducting MgB$_2$  tip
\cite{Giubileo2}. In such a SIS geometry the spectroscopic
features are enhanced by the convolution of the DOS of two
electrodes. One can clearly see a ZB peak characteristic for SIS
junctions, much stronger quasiparticle peaks appearing at $\pm$6
meV and not at $\pm$3 meV as in SIN spectra (Fig.1). Remarkably,
some additional features, humps and dips are seen on the tails of
the quasiparticle peaks. The best fit using BCS DOS is plotted as
a thin solid line, the thick one representing the best fit using
McMillan approach (2). It is clear that the thick line follows all
measured spectroscopic features in finer details. Such an
excellent agreement indicates that one should include into the
realistic model of the superconductivity in MgB$_2$ a contribution
of the quasiparticle inter-band scattering. The values for
electron-phonon coupling constants should be probably
reconsidered.

We see that the use of both models result in a significant drop of
the superconducting gap energy at 0.6 T. This finding is in
qualitative agreement with the theoretical prediction
\cite{Koshelev}. The relative change of the superconducting gap is
larger within Mac Millan model (2) since there a significant part
of the gapped states originates from the quasiparticle scattering
from the $\sigma$-band, a process which is field independent. Both
models suggest that at the field of 0.6T the contribution to the
superconductivity from the electron-electron interaction via
phonons in the $\pi$-band itself is not efficient anymore. At
higher fields the gap in the tunneling DOS survives due to the
phonon exchange with $\sigma$-band \cite{Suhl} or due to the both,
phonon exchange and quasiparticle inter-band scattering (Mac
Millan model (2) ).

\begin{figure}
\includegraphics[width=8.5cm]{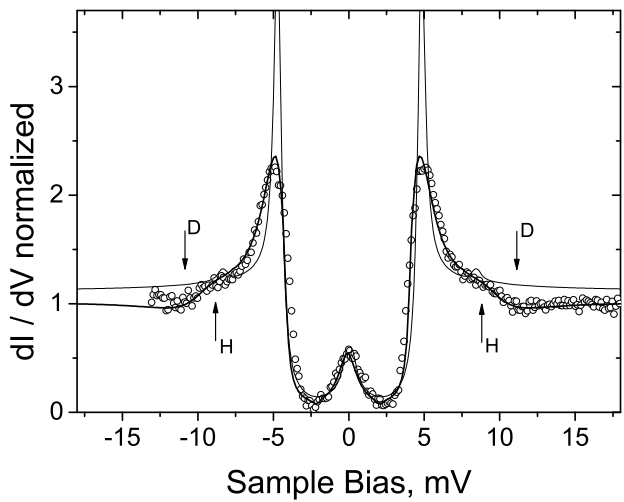}
\caption{\label{fig:epsart} Circles: Zero field S-I-S tunneling
spectrum. Thin solid line: fit using (1) with $\Delta_{\sigma}$ =
6.6 meV (relative weight 2\%), $\Delta_{\pi}$(98\%) = 2.35 meV,
$\Gamma_{Dynes}$ = 0.1 meV, T = 12 K; thick solid line: fit using
(2) with $\Delta_{\sigma}^0$ = 7.4 meV, $\Delta_{\pi}^0$ = 1.0
meV, $\Gamma_{\sigma}$ = 2.3 meV, $\Gamma_{\pi}$ = 2.5 meV,
$\Gamma_{Dynes}$ = 0.12 meV, T = 12 K. Arrows indicate the
positions of humps (H) and dips (D) perfectly reproduced by the
Mac Millan approach.}
\end{figure}

We note here that while equation (1) is often used to describe the
quasiparticle spectrum in the magnetic field \cite{Gonnelli}, its
application is not rigorously justified from a theoretical point
of view. Indeed, (1) and (2) initially describe the quasiparticle
spectrum in zero field. Even the introduction of $\Gamma_{Dynes}$
to consider  the pair breaking in the magnetic field is not really
satisfactory as it is correct only at B$\sim$B$_{c2}$. We decided
to use such a procedure in order to clarify the physical meaning
of the field dynamics we experimentally observed, in the lack of a
realistic theory describing the DOS in the magnetic field.

In conclusion, in this paper we studied the superconducting
$\pi$-band tunneling DOS in single crystals of MgB$_2$  in
magnetic field, and we succeeded to observe the superconducting
gap up to 2 T. The evolution of the DOS is characterized by two
distinct regimes separated by a crossover region. Our results
indicate a rapid suppression of the intrinsic term in $\pi$-band
superconductivity for $0$ T $< B < 0.4$ T. At high fields ($0.7$ T
$< B < 2 $ T) the superconductivity in the $\pi$-band survives
uniquely due to the coupling to the $\sigma$-band. The shape of
tunneling spectra suggests an important role played by the
quasiparticle inter-band scattering. We think an additional
theoretical effort is needed to consider properly the inter-band
coupling in the realistic model of the two-band superconductivity
in MgB$_2$.

The authors thank A.A. Golubov for useful discussions. This work
has been supported by Italian MIUR project "$\it{Rientro \, dei \,
Cervelli}$" and by the French University Paris 6 PPF project.


\end{document}